\documentclass[11pt,reqno,twoside]{article}
\usepackage{amsmath,amssymb,theorem}
\theorembodyfont{\itshape}
\newtheorem{thm}{Theorem}[section]
\newtheorem{lem}[thm]{Lemma}
\newtheorem{prop}[thm]{Proposition}
{\theorembodyfont{\upshape}
\newtheorem{define}[thm]{Definition}
\newtheorem{rem}[thm]{Remark}
\newtheorem{ass}[thm]{Assumption}}
\newtheorem{cor}[thm]{Corollary}

\newcommand{\Proof}[1][]{\noindent{\itshape Proof#1: }}
\newcommand{\EndProof}{~$\Box$\bigskip}

\makeatletter
    
    \@addtoreset{equation}{section}
\makeatother

\def\hat{\widehat}

\def\mbR{{\mathbb R}}
\def\mbC{{\mathbb C}}
\def\mbE{{\mathbb E}}
\def\mbZ{{\mathbb Z}}

\def\mcC{\mathcal{C}}
\def\mcD{\mathcal{D}}

\def\mcF{\mathcal{F}}
\def\mcG{\mathcal{G}}

\def\mcN{\mathcal{N}}
\def\mcO{\mathcal{O}}
\def\mcR{\mathcal{R}}

\def\ua{\uparrow}
\def\da{\downarrow}

\def\ul{\underline}
\def\sm{\setminus}
\def \sbs{\subset}
\def\pd{\prod}
\def\sps{\supset}
\def\ssbs{\subset\subset}
\def\ptl{\partial}
\def\es{\emptyset}
\def\fa{\forall}
\def\Tr{\text{Tr}}

\def\d{\delta}    
\def\a{\alpha}           \def\et{\eta}       \def\d{\delta}
\def\D{\Delta}     \def\ph{\varphi}       \def\Ph{\Phi}
  \def\g{\gamma}      \def\G{\Gamma}     
\def\l{\lambda}   \def\L{\Lambda}     \def\m{\mu}        
\def\r{\rho}               
\def\p{\pi}              \def\s{\sigma}     
          \def\ch{\chi}
\def\x{\xi}       
\def\z{\zeta}           

\begin{document}
\title{Gibbsianness of Fermion Random Point Fields}
\author{ Hyun Jae Yoo\footnotemark[1] }
\date{}
  \maketitle

\begin{abstract}
We  consider fermion (or determinantal) random point fields on
Euclidean space $\mbR^d$. Given a bounded, translation invariant,
and positive definite integral operator $J$ on $L^2(\mbR^d)$, we
introduce a determinantal interaction for a system of particles
moving on $\mbR^d$ as follows: the $n$ points located at
$x_1,\cdots,x_n\in \mbR^d$ have the potential energy given by
$$
U^{(J)}(x_1,\cdots,x_n):=-\log\det(j(x_i-x_j))_{1\le i,j\le n},
$$
where $j(x-y)$ is the integral kernel function of the operator
$J$. We show that the Gibbsian specification for this interaction
is well-defined. When $J$ is of finite range in addition, and for
$d\ge 2$ if the intensity is small enough, we show that the
fermion random point field corresponding to the operator
$J(I+J)^{-1}$ is a Gibbs measure admitted to the specification.
\end{abstract}
{\bf Running head}. {Gibbsianness of FRPF's}

\footnotetext[1]{University College , Yonsei University, 134
Shinchon-dong, Seodaemoon-gu, Seoul 120-749, Korea. E-mail:
yoohj@yonsei.ac.kr}

\section{Introduction}
Fermion (or determinantal) random point fields (FRPF's hereafter)
are probability measures on the configuration space of particles
(moving on discrete or continuum spaces) whose correlation
functions are determined by determinants of matrices. See Section
2 for the definition. In many literature FRPF's are investigated;
the problem of existence, basic properties, ergodicity (for
translationally invariant case), stochastic domination, and
connection to other physical problems  have been studied \cite[and
references therein]{BOO, BO, GY, L, LS, M, ST1, ST2, So}.

The aim of this paper is to investigate the Gibbsianness of FRPF's
on continuum spaces and also to construct the suitable
interactions. The Gibbsianness of FRPF's on discrete spaces was
first shown in \cite{ST2} and then in \cite{SY} in different ways
for suitable FRPF's. Recently, Georgii and the present author
studied the conditional intensity of FRPF's on continuum spaces
and considered the Gibbsianness in a different way \cite{GY}. In
this paper, we particularly focus on the Hamiltonian and Gibbsian
specification to which certain FRPF's are admitted. It provides us
with a new view point for FRPF's. In addition, it has also other
merits. First, there are not so many non-trivial examples of
interactions for particle systems moving on continuum spaces for
which the equilibrium measures (Gibbs measures) are proved to
exist. The typical examples are the superstable interactions
introduced by Ruelle \cite{R1, R2}. The Gibbsianness of FRPF's
thus gives rise to another example of interactions. One more
benefit comes from its applicability when one wants to construct
some dynamics of particles for which given FRPF's are invariant.
In \cite{SY} we have constructed the Glauber dynamics on the
discrete space leaving a given FRPF invariant. In \cite{Y}, we
have constructed the Dirichlet forms and the associated diffusion
processes on the configuration space of particles moving on
continuum spaces for which certain FRPF's are invariant.

We briefly summarize  the contents of this paper. Let $J$ be a
bounded, positive definite integral operator on $L^2(\mbR^d)$.
Suppose that its kernel function $J(x,y)$, $x,y\in \mbR^d$, is
bounded, continuous, and translation invariant, i.e., there is a
bounded and continuous function $j(x)$ of positive type \cite{RS}
such that
\begin{equation}
J(x,y)=j(x-y).
\end{equation}
By using it we define the potential energy
$U^{(J)}(x_1,\cdots,x_n)$ of $n$ particles located at $x_1,\cdots,
x_n$ by
\begin{equation}
U^{(J)}(x_1,\cdots,x_n):=-\log\det(j(x_i-x_j))_{1\le i,j\le n}.
\end{equation}
We show that the Gibbsian specification for the interaction is
well-defined (Proposition \ref{prop:specification}). Furthermore,
we show that if $j(x)$ is of finite range, i.e., there is some
$R>0$ such that $j(x)=0$ if $|x|\ge R$ and, for $d\ge 2$, if the
intensity $j(0)$ is sufficiently small, the FRPF corresponding to
the operator $J(I+J)^{-1}$ is a Gibbs measure for the
specification (Theorem  \ref{thm:Gibbsian}).

This paper is organized as follows. In Section  \ref{sec:pre}, we
review the definition of FRPF's with basic properties. In Section
\ref{sec:res}, we define a Gibbsian specification and state the
main results. Section \ref{sec:pr} is devoted to the proofs. In
Section \ref{sec:conc}, we discuss some possible improvements. In
Appendix, we give examples of bounded and continuous functions of
positive type which have finite ranges.

\vspace{0cm plus 4cm}

\section{Preliminaries}\label{sec:pre}
\subsection{FRPF's on Continuum Spaces}
In this subsection we briefly recall the definition of FRPF's. For
a more complete survey on this field, we refer to the articles
\cite{BOO, BO, GY, L, LS, M, ST1, ST2, So}. The state space for
FRPF's may be a very general separable Hausdorff space but in this
paper we fix it to be $\mbR^d$. It is understood as a one particle
space.

We denote by $\mcN$ the space of locally finite, integer-valued
Radon measures on $\mbR^d$, equipped with the vague topology. We
notice that an element (called a configuration) $\x\in \mcN$ is
expressible as
\begin{equation}\label{eq:configuration}
\x=\sum_ik_i\d_{x_i},
\end{equation}
where each $k_i$ is a positive integer and $\d_{x_i}$ is a Dirac
measure, and distinct points $\{x_i\}$ form a countable set with
at most  finitely many $x_i$'s in any bounded Borel subset of
$\mbR^d$. We recall that $\mcN$ is a Polish space, i.e., $\mcN$
can be given a metric so that it becomes a complete separable
metric space. Moreover, the induced topology from that metric is
equivalent to the vague topology \cite[Corollary 7.1.IV and
section A2.6]{DV}. The Borel $\s$-aglebra $\mcF$ on $\mcN$ is
equal to the smallest $\s$-algebra with respect to which the
mappings
\begin{equation}
\x\mapsto N_\L(\x):=\x(\L)
\end{equation}
are measurable for any bounded Borel subset $\L\sbs\mbR^d$
\cite[Corollary 7.1.VI]{DV}. For each Borel subset $\D\sbs\mbR^d$,
we let $\mcF_\D$ the $\s$-algebra on $\mcN$ generated by $N_\L$'s
where $\L$ runs over all bounded Borel subsets of $\D$.

In the sequel a measurable subset of $\mcN$ will play a central
role. Recall that $\x\in \mcN$ is called simple if all the $k_i$'s
are $1$ in the representation (\ref{eq:configuration}). The space
of all simple measures is denoted by $\G$, which is a measurable
subset of $\mcN$ \cite[Proposition 7.1.III]{DV}. We will denote by
$\mcF^{(\G)}$ resp. $\mcF_\D^{(\G)}$ the $\s$-algebras
$\{A\cap\G:\,A\in \mcF\}$ resp. $\{A\cap\G:\,A\in \mcF_\D\}$ in
$\G$.

By a {\it random point field} (abbreviated RPF) we mean a triple
$(\mcN,\mcF,\m)$ where $\m$ is a probability measure on $\mcF$.
For simplicity we call such a measure $\m$ itself as a RPF.
\begin{define}\label{def:correlation}
A locally integrable function $\r_n:(\mbR^d)^n\to \mbR_+$ is
called the $n$-point correlation function of a RPF $\m$ if for any
disjoint bounded Borel subsets $\L_1,\cdots,\L_m$ of $\mbR^d$ and
$k_i\in \mbZ_+$, $i=1,\cdots,m$, $\sum_{i=1}^mk_i=n$, the
following identity holds:
\begin{equation}
\mbE_\m
\pd_{i=1}^m\frac{(N_{\L_i})!}{(N_{\L_i}-k_i)!}=\int_{\L_1^{k_1}\times\cdots\times
\L_m^{k_m}} \r_n(x_1,\cdots,x_n)dx_1\cdots dx_n,
\end{equation}
where $\mbE_\m$ denotes the expectation w.r.t. $\m$ and $dx$ is
the Lebesgue measure on $\mbR^d$.
\end{define}
\begin{define} A RPF is called fermion (or determinantal) if its $n$-point correlation
functions are given by
\begin{equation}\label{eq:correlation_function}
\r_n(x_1,\cdots,x_n)=\det(K(x_i,x_j))_{1\le i,j\le n},
\end{equation}
where $K(x,y)$, $x,y\in \mbR^d$, denotes the integral kernel
function of an integral operator $K$ on $L^2(\mbR^d)$.
\end{define}
For the existence   of FRPF's we state the following theorem from
\cite{So} (see also \cite{M, ST1}). We denote by $I$ the identity
operator on $L^2(\mbR^d)$.
\begin{thm}\label{thm:frpf}  Hermitian locally trace class operator $K$ on $L^2(\mbR^d)$
defines a FRPF if and only if $0\le K\le I$. If the corresponding
FRPF exists it is unique.
\end{thm}
In this paper we restrict ourselves to the Hermitian operators for
the defining operator $K$, but there are examples of FRPF's with
non-Hermitian operators \cite{BOO}. We notice that from the
determinantal nature of the correlation functions in
(\ref{eq:correlation_function}), FRPF's are in fact measures on
$(\G,\mcF^{(\G)})$. Below we summarize some basic properties of
FRPF's.

\subsection{Basic Properties of FRPF's}
First we remark that any FRPF has a system of density
distributions. Recall that the density distributions, or called
the Janossy densities \cite{DV}, of a RPF $\m$ are the measurable
functions $(\s_\L^m)$, where $m\in \mbZ_+$ and $\L$ runs over all
bounded Borel subsets of $\mbR^d$, that satisfy following
properties \cite{R2}:

(a) (Symmetry)
$\s_\L^m(x_{i_1},\cdots,x_{i_m})=\s_\L^m(x_1,\cdots,x_m)$ for
every permutation $(1,\cdots,m)$ $\to (i_1,\cdots,i_m)$.

(b) (Normalization)
\begin{equation}
\sum_{m=0}^\infty\frac1{m!}\int_{\L^m}\s_\L^m(x_1,\cdots,x_m)dx_1\cdots
dx_m=1.
\end{equation}

(c) (Compatibility) If $\L\sbs\D$, then
\begin{equation}
\s_\L^m(x_1,\cdots,x_m)=\sum_{n=0}^\infty\frac1{n!}\int_{(\D\sm\L)^n}\s_{\D}^{m+n}
(x_1,\cdots,x_{m+n})dx_{m+1}\cdots dx_{m+n}.
\end{equation}
The relation between $\m$ and $(\s_\L^m)$ is given by the
following properties: if $f:\mcN\to \mbR$ is any measurable local
(cylindrical) function, say $\L$-local, then
\begin{equation}
\int
f(\x)d\m(\x)=\sum_{m=0}^\infty\frac1{m!}\int_{\L^m}f(x_1,\cdots,x_m)\s_\L^m(x_1,\cdots,x_m)
dx_1\cdots dx_m.
\end{equation}
Moreover, the correlation functions of $\m$ are then recovered
from $(\s_\L^m)$ by the following relation:
\begin{equation}
\r_n(x_1,\cdots,x_n)=
\sum_{m=0}^\infty\frac1{m!}\int_{\L^m}\s_{\L}^{n+m}(x_1,\cdots,x_{n+m})dx_{n+1}\cdots
dx_{n+m}
\end{equation}
for $x_1,\cdots,x_n\in \L$.

For each bounded Borel subset $\L\sbs\mbR^d$, we denote by $P_\L$
the projection from $L^2(\mbR^d)$ onto $L^2(\L)$. Let $\m$ be any
FRPF corresponding to an operator $K$ (see Theorem
\ref{thm:frpf}). Let $K_\L:=P_\L KP_\L$ be the restriction of $K$
to $L^2(\L)$ and $K_\L(x,y)$ its kernel function. That is,
$K_\L(x,y)=1_\L(x)K(x,y)1_\L(y)$ with $1_\L$ being the
characteristic function on the set $\L$. The density functions of
$\m$ are given by \cite{ST1, So}
\begin{equation}\label{eq:density_function}
\s_\L^m(x_1,\cdots,x_m)=\det(I-K_\L)\det(J_{[\L]}(x_i,x_j))_{1\le
i,j\le m}, \quad x_i\in \L,\,\,i=1,\cdots,m,
\end{equation}
where $\det(I-K_\L)$ is a Fredholm determinant \cite{Sim} and
$J_{[\L]}:=K_\L(I-K_\L)^{-1}$. In the following remark we gather
some basic facts about the density distributions for FRPF's.
\begin{rem} \label{rem:basic-properties}
(i) The formula (\ref{eq:density_function}) is well-defined even
in the case $1\in \text{spec}\,K_\L$, the spectrum of $K_\L$. See
\cite{So}.

(ii) We recall that given a trace class operator $T$ on
$L^2(\mbR^d)$ the Fredholm determinant of $I+T$ is given by
\begin{equation}
\det(I+T)=\sum_{n=0}^\infty \Tr(\wedge^nT),
\end{equation}
where $\wedge^nT$ is the $n$-th exterior product of $T$ and the
following estimate holds:
\begin{equation}\label{eq:exterior}
 \|\wedge^nT \|_1\le \frac1{n!} \|T \|_1^n,
\end{equation}
where $ \|\cdot \|_1$ is the trace norm. For the density
distributions of FRPF's we have the following relation:
\begin{eqnarray}\label{eq:number_density}
\mbE(1_{\{N_\L=m\}})&=&\frac1{m!}\int_{\L^m}\s_\L^m(x_1,\cdots,x_m)dx_1\cdots dx_m\nonumber\\
&=&\det(I-K_\L)\frac1{m!}\int \det(J_{[\L]}(x_i,x_j))_{1\le i,j\le m}dx_1\cdots dx_m\nonumber\\
&=&\det (I-K_\L)\Tr(\wedge^m(J_{[\L]})).
\end{eqnarray}
The last equality follows from \cite[Theorem 3.10]{Sim} if the
kernel function $J_{[\L]}(x,y)$ is continuous. The general case
follows from the argument of \cite{So} (see the equation (1.26) of
\cite{So}).

(iii) If $\m$  is a FRPF corresponding to an integral operator
$K$, using the expression (\ref{eq:density_function})  we have the
following Laplace transform of $\m$ (cf. \cite{ST1}): for $f\in
C_0(\mbR^d)$,
\begin{equation}\label{eq:Laplace_transform}
\int \exp(-<f,\x>)d\m(\x)=\det(I-K\psi_f),
\end{equation}
where $<f,\x>:=\sum_{x_i\in \x}f(x_i)$ and
$\psi_f(x)=1-\exp(-f(x))$ and $K\psi_f$ is the product of $K$ and
the multiplication operator with function $\psi_f$, and the
determinant is a Fredholm determinant.
\end{rem}

\vspace{0cm plus 4cm}

\section{Results}\label{sec:res}
\subsection{Determinantal Potentials and Gibbsian Specifications}
In this subsection we introduce a particle system with an
interaction which is given by determinants of matrices. The matrix
components are given by the kernel function of an integral
operator on $L^2(\mbR^d)$. We consider bounded linear operators
$J$ on $L^2(\mbR^d)$  constructed in the following way:
\begin{ass}\label{ass:assumption1} The operator $J$ is defined as an integral operator on
$L^2(\mbR^d)$ with integral kernel function
\begin{equation}\label{kernel-function}
J(x,y):=j(x-y),
\end{equation}
where $j(\cdot)\in L^1(\mbR^d)$ is a (inverse) Fourier transform
of a finite measure $d\r$ on $\mbR^d$:
\begin{equation}\label{eq:Fourier-transform}
j(x):=(2\p)^{-d}\int_{\mbR^d}e^{ix\cdot t}d\r(t).
\end{equation}
\end{ass}
By Bochner's theorem and Young's inequality \cite{RS}, the
operators $J$ in Assumption \ref{ass:assumption1} are bounded,
positive definite linear operators on $L^2(\mbR^d)$.

We fix an operator $J$ satisfying the conditions in Assumption
\ref{ass:assumption1}. For each integer $n\ge 0$ and
$x_1,\cdots,x_n\in \mbR^d$, the potential energy
$U^{(J)}(x_1,\cdots,x_n)$ of $n$ particles located at $x_1,\cdots,
x_n$ is defined by
\begin{equation}\label{eq:potential}
U^{(J)}(x_1,\cdots,x_n):=-\log\det(j(x_i-x_j))_{1\le i,j\le n}.
\end{equation}
Since the matrix $(j(x_i-x_j))_{1\le i,j\le n}$ as an operator on
$\mbC^n$ is positive definite the function $\det(j(x_i-x_j))_{1\le
i,j\le n}$ is nonnegative. For a convenience we set $-\log 0\equiv
+\infty$. Then $U^{(J)}(x_1,\cdots,x_n)$ is well-defined with
values in $\mbR\cup \{+\infty\}$ (if $n=0$, we set $U^{(J)}=0$),
in particular if some of two points $x_i$ and $x_j$ in
$(x_i,\cdots, x_n)$ are the same, then
$U^{(J)}(x_1,\cdots,x_n)=+\infty$, i.e., under this interaction,
two or more particles can not share a single point. It is obvious
that $U^{(J)}$ is translation invariant. Moreover, we notice that
\begin{equation}\label{eq:consistency}
\det \left(\begin{array}{cc}
A&B\\
B^*&C\end{array}\right)=0 \text{ whenever } \det A=0 \text{ or
}\det C=0
\end{equation}
for any positive definite matrices $\left(\begin{array}{cc}
A&B\\
B^*&C\end{array}\right)$ ($B^*$ denotes the adjoint matrix of
$B$). This follows from the following Fischer's inequality:
\begin{equation}\label{eq:determinant-inequality}
\det \left(\begin{array}{cc}
A&B\\
B^*&C\end{array}\right)\le \det A\,\det C.
\end{equation}
Thus, $U^{(J)}$ is uniquely decomposed as
\begin{equation}\label{eq:interaction}
U^{(J)}(x_1,\cdots,x_n)=\sum_k\sum_{1\le i_1<\cdots<i_k\le
n}\Ph_k^{(J)}(x_{i_1},\cdots,x_{i_k})
\end{equation}
with some functions $\Ph_k^{(J)}$. Notice that the $k$-body
potential $\Ph_k^{(J)}$ is invariant under permutation of its $k$
arguments and under translations in $\mbR^d$. We call the sequence
$(\Ph_k^{(J)})_{k\ge 1}$ of $k$-body potentials the interaction
 determined by the operator $J$ \cite{R1}.

We now construct a Gibbsian specification for the interaction
determined by $J$. For conveniences, we introduce the following
notations.

\medskip

 \noindent {\bf Notations}: Each element $\x\in
\mcN$ will also be understood as a finite or countably infinite
sequence $\x=(x_1,x_2,\cdots)$ in $\mbR^d$ determined by the
support of the measure $\x$ (see \eqref{eq:configuration}). Of
course, some of the components are repeated in general, but when
$\x\in \G$, all the components are distinct. Any set of finite
points $(x_1,\cdots,x_n)\in (\mbR^d)^n$ is denoted by $\ul x_n$.
For any $ \L\sbs\mbR^d$, $\x_\L$ represents a configuration on
$\L$ or a restriction of a configuration $\x\in \mcN$ to the
region $\L$, i.e., $\x_\L=\x\cap\L$. $\mcN_\L$ denotes the set of
all configurations $\x$ such that $\x=\x_\L$ and set
$\G_\L:=\mcN_\L\cap \G$. If $\L_1$ and $\L_2$ are two disjoint
subsets of $\mbR^d$, by $\x_{\L_1}\z_{\L_2}$ we denote a
configuration in $\mcN_{\L_1\cup\L_2}$ which coincides with
$\x_{\L_1}$ on $\L_1$ and with $\z_{\L_2}$ on $\L_2$. Any bounded
Borel subset $\L\sbs\mbR^d$ is denoted by $\L\ssbs\mbR^d$. For any
function $A(\cdot,\cdot):\mbR^d\times\mbR^d\to \mbC$ and $\ul
x_n=(x_1,\cdots,x_n)\in (\mbR^d)^n$, $A(\ul x_n,\ul x_n)$ denotes
the finite matrix
\begin{equation}
A(\ul x_n,\ul x_n):=(A(x_i,x_j))_{1\le i,j,\le n}.
\end{equation}
Finally, for any $\L\ssbs\mbR^d$ and a $\L$-local function $f$ we
simplify the integration
\begin{equation}
\sum_{n\ge 0}\frac{1}{n!}\int_{\L ^n}d\ul x_n f(\ul x_n)=:\oint_\L
d\z _\L  f(\z _\L ), \label{eq:integral}
\end{equation}
where $d\ul x_n$ denotes the Lebesgue measure on $(\mbR^d)^n$.

We now consider the energy of a particle configuration in a bounded region with a given boundary condition. We
will need to refine the boundary particles that would interact with some particles inside the region. We say
that the system has an interaction range $R\in (0,+\infty]$ defined by
\begin{equation}\label{eq:interaction-range}
R:=\inf\{R'\in \mbR:\,j(x)=0 \text{ whenever } |x|\ge R'\}.
\end{equation}
(We consider only the case $R>0$.) Given a region $\L\ssbs\mbR^d$ and a configuration
$\x=(x_i)_{i=1,2,\cdots}\in \mcN$, we define a subset $\x_{\ptl\L}\sbs\x_{\L^c}$ of boundary particles  that
interact with the particles inside $\L$ as follows. In the case $R= +\infty$, we let $\x_{\ptl\L}\equiv
\x_{\L^c}$. In the case $R<\infty$, we say that a particle $x_i\in \x_{\L^c}$ interacts with particles in the
region $\L$ if there is a finite sequence $(x_{j_1},\cdots,x_{j_k})\sbs\x_{\L^c}$ and a point $x_{j_0}\in \L$
such that $x_{j_k}=x_i$ and $|x_{j_l}-x_{j_{l-1}}|<R$ for $l=1,\cdots,k$. We define
\begin{equation}\label{eq:true-boundary-particles}
\x_{\ptl\L}:=\{x_i\in \x_{\L^c}:\, x_i\text{ interacts with
particles inside }\L\}.
\end{equation}

From the decomposition (\ref{eq:interaction}) we see that for any
$\L_1,\L_2\ssbs\mbR^d$ with $\L_1\cap\L_2=\es$, and finite
configurations $\x_{\L_1}$ and $\x_{\L_2}$, the mutual potential
energy $W^{(J)}(\x_{\L_1};\x_{\L_2})$ is well-defined to satisfy
\begin{equation}\label{eq:mutual_potential}
U^{(J)}(\x_{\L_1\cup\L_2})=U^{(J)}(\x_{\L_1})+U^{(J)}(\x_{\L_2})+W^{(J)}(\x_{\L_1};\x_{\L_2})
\end{equation}
if $U^{(J)}(\x_{\L_1\cup\L_2})<\infty$, and
$W^{(J)}(\x_{\L_1};\x_{\L_2})=\infty$ if
$U^{(J)}(\x_{\L_1\cup\L_2})=\infty$. Now for each $\L\ssbs\mbR^d$,
$\z_\L\in \mcN_\L$, and $\x\in \mcN$, we define the energy of the
particle configuration $\z_\L$ on $\L$ with boundary condition
$\x$ by
\begin{equation}\label{eq:energy-with-b.c.}
H_\L^{(J)}(\z_\L;\x):=\lim_{\D\ua\mbR^d}[U^{(J)}(\z_\L)+W^{(J)}(\z_{\L};\hat\x_{\D\sm\L})],
\end{equation}
whenever the limit exists. Here $\hat\x_{\D\sm\L}$ is defined by
\begin{equation}
\hat\x_{\D\sm\L}:=\x_{\D\sm\L}\cap \x_{\ptl\L}.
\end{equation}
In Lemma \ref{lem:specification-density} below we show that
$H_\L^{(J)}(\z_\L;\x)$ does exist for all $\z_\L\in \G_\L$ and
\lq\lq physically possible\rq\rq  configurations $\x\in \G$. For
that purpose we introduce the following events. For each
$\L\ssbs\mbR^d$, define a subset $\mcR_\L\in \mcF_{\L^c}$, which
will represent the \lq\lq possible\rq\rq  event in $\mcF_{\L^c}$
(see \cite[page 16]{Pr}), as follows:
\begin{equation}\label{eq:possible-event}
\mcR_\L:=\{\x\in \mcN:\det (J(\hat\x_\D,\hat\x_\D))\neq 0,\quad
\fa\,\D\ssbs\L^c\},
\end{equation}
where as before $\hat\x_\D:=\x_\D\cap\x_{\ptl\L}$.
\begin{lem}\label{lem:specification-density}
Suppose that $J$ is an integral operator on $L^2(\mbR^d)$
satisfying the conditions in Assumption \ref{ass:assumption1}.
Then for any $\L\ssbs\mbR^d$, $\z_\L\in \G_\L$, and $\x\in
\mcR_\L$, the function $H_\L^{(J)}(\z_\L;\x)$ in
(\ref{eq:energy-with-b.c.}) is well-defined.
\end{lem}

We are now ready to define the Gibbsian specification. For a
convenience we extend the function $H_\L^{(J)}(\z_\L;\x)$ to the
whole $\z_\L\in \mcN_\L$ and $\x\in \mcN$. We set
\begin{equation}
H_\L^{(J)}(\z_\L;\x)\equiv +\infty \text{ unless }\z_\L\in \G_\L
\text{ and } \x\in \mcR_\L.
\end{equation}
For each $\L\ssbs\mbR^d$, $\z_\L\in \mcN_\L$, and $\x\in \mcN$, we
define the function
\begin{equation}\label{eq:specification-density}
\g_\L^{(J)}(\z_\L;\x):=\begin{cases}\frac1{Z_\L^{(J)}(\x)}
\exp[-H_\L^{(J)}(\z_\L;\x)], \text{ if }\z_\L\in \G_\L
\text{ and } \x\in \mcR_\L,\\
0,\,\,\text{otherwise}.\end{cases}
\end{equation}
In the above $Z_\L^{(J)}(\x)$ is the partition function, i.e., a
normalization constant defined by
\begin{equation}\label{eq:normalization}
Z_\L^{(J)}(\x):=\sum_{n\ge 0}\frac1{n!}\int_{\L^n}d\ul
x_n\exp[-H_\L^{(J)}(\ul x_n;\x)].
\end{equation}
We let $J_\L$ denote the restriction of the operator $J$ to
$L^2(\L)$:
\begin{equation}
J_\L:=P_\L JP_\L.
\end{equation}
It turns out that for any $\x\in \mcR_\L$, $Z_\L^{(J)}(\x)$ is a
finite number satisfying (see (\ref{eq:partition-function-bound}))
\begin{equation}
1\le Z_\L^{(J)}(\x)\le \det(I+J_\L),
\end{equation}
where $\det(I+J_\L)$ is the Fredholm determinant of the operator
$I+J_\L$. Now for any bounded measurable function $f:\,\mcN\to
\mbR$, $\L\ssbs\mbR^d$, and $\x\in \mcN$, we define
\begin{equation}\label{eq:specification}
\g_\L^{(J)}(f|\x):=\oint_\L d\z_\L
\g_\L^{(J)}(\z_\L;\x)f(\z_\L\x_{\L^c}).
\end{equation}
We will prove that the system
$(\g_\L^{(J)}(\cdot|\cdot))_{\L\ssbs\mbR^d}$ defines a
specification.
\subsection{Gibbsianness of FRPF's}
We start by summarizing the construction in the last subsection.
\begin{prop}\label{prop:specification}
Suppose that $J$ satisfies the conditions in Assumption
\ref{ass:assumption1}. Then the system
$(\g_\L^{(J)}(\cdot|\cdot))_{\L\ssbs\mbR^d}$ given in
(\ref{eq:specification}) defines a specification with respect to
${\mcR}:=({\mcR}_\L)_{\L\ssbs\mbR^d}$ (see \cite[page 16]{Pr}).
\end{prop}
The main purpose of this paper is to characterize the Gibbs
measures admitted to the specification
$(\g_\L^{(J)})_{\L\ssbs\mbR^d}$ in the above. Recall that a
probability measure $\m$ on $(\mcN,\mcF)$ is said to be admitted
to $(\g_\L^{(J)})_{\L\ssbs\mbR^d}$, or to satisfy the DLR
equations (see \cite{G, Pr}) if for any $\L\ssbs\mbR^d$ and
bounded measurable function $f:\mcN\to \mbR$,
\begin{equation}\label{eq:DLR-equation}
\int d\m(\x)f(\x)=\int d\m(\x)\g_\L^{(J)}(f|\x).
\end{equation}

Suppose that $J$ is an integral operator as in Assumption
\ref{ass:assumption1}. We define
\begin{equation}\label{eq:K-operator}
K^{(J)}:=J(I+J)^{-1}.
\end{equation}
$K^{(J)}$ then is a locally trace class operator and satisfies
$0\le K^{(J)}< I$. Therefore by Theorem \ref{thm:frpf} defines a
FRPF which we denote by $\m^{(J)}$. We conjecture that $\m^{(J)}$
is a Gibbs measure for the specification in Proposition
\ref{prop:specification}. Unfortunately, however, we couldn't
completely prove it. We impose further conditions on the operator
$J$:
\begin{ass}\label{ass:assumption2}
In addition to the conditions in Assumption \ref{ass:assumption1},
we assume that the finite measure $d\r(t)$ in
(\ref{eq:Fourier-transform}) has a density:
$d\r(t)=\hat{\ph}(t)dt$, and $j(\cdot)$ is of finite range, i.e.,
there exists $0<R<\infty$ such that
\begin{equation}\label{eq:finite-range}
j(x)=0 \text{ if }|x|\ge R.
\end{equation}
\end{ass}
In the Appendix, we provide with some examples of $j(\cdot)$ in
Assumption \ref{ass:assumption2}. We call the finite number
$J(0,0)\equiv j(0)$ the intensity of the system (in \cite[page
112]{M}, the terminology \lq\lq intensity\rq\rq was used for the
quantity $K^{(J)}(0,0)$, but the two are similar in nature). The
following is a main result of this paper:
\begin{thm}\label{thm:Gibbsian}
Suppose that $J$ is an integral operator on $L^2(\mbR^d)$
satisfying the conditions in Assumption \ref{ass:assumption2}. For
$d\ge 2$, assume further that the intensity $j(0)$ is small
enough. Then the corresponding FRPF $\m^{(J)}$ is a Gibbs measure
admitted to the specification $(\g_\L^{(J)})_{\L\ssbs\mbR^d}$ in
Proposition \ref{prop:specification}.
\end{thm}
We also introduce the activity of the system. We recall \cite{R1}
that by an activity $z>0$ of the system we mean that for any
$\L\ssbs\mbR^d$, a grand canonical ensemble on $\sum_{n\ge
0}^\infty\L^n$ is a measure with restriction to $\L^n$ given by
\begin{equation}
\frac{z^n}{n!}\exp[-U^{(J)}(x_1,\cdots,x_n)]dx_1\cdots x_n.
\end{equation}
Analogously, for each $z>0$ we define a new specification
$(\g_\L^{(J;z)})_{\L\ssbs\mbR^d}$ by multiplying $z^{|\z_\L|}$,
$|\z_\L|$ being  the cardinality of $\z_\L$, in front of
$\exp[-H_\L^{(J)}(\z_\L;\x)]$ in (\ref{eq:specification-density})
and suitably re-defining the partition function $Z_\L^{(J;z)}(\x)$
as
\begin{equation}\label{eq:new-normalization}
Z_\L^{(J;z)}(\x):=\sum_{n\ge 0}\frac{z^n}{n!}\int_{\L^n}d\ul
x_n\exp[-H_\L^{(J)}(\ul x_n;\x)].
\end{equation}
We say that the system has an activity $z$.
\begin{cor}\label{cor:small-activity-case}
Assume that $J$ is an integral operator satisfying the conditions
in Assumption \ref{ass:assumption2}. If the activity $z>0$ of the
system is sufficiently small then the FRPF $\m^{(zJ)}$ is a Gibbs
measure for the specification $(\g_\L^{(J;z)})_{\L\ssbs\mbR^d}$.
\end{cor}
\Proof It is easily seen from (\ref{eq:potential}) and
(\ref{eq:mutual_potential})-(\ref{eq:energy-with-b.c.}) that
\begin{equation}
H_\L^{(zJ)}(\z_\L;\x)=-\log z^{|\z_\L|}+ H_\L^{(J)}(\z_\L;\x).
\end{equation}
(See (\ref{eq:hamiltonian-reexpression}).) Thus the scaling
property $\g_\L^{(J;z)}=\g_\L^{(zJ)}$ holds, i.e., the
specification $(\g_\L^{(J;z)})_{\L\ssbs\mbR^d}$ is the same as
$(\g_\L^{(zJ)})_{\L\ssbs\mbR^d}$. Therefore, smallness of the
activity implies smallness of the intensity $zj(0)$ of the system
distributed by $\m^{(zJ)}$. The conclusion follows from Theorem
\ref{thm:Gibbsian}. \EndProof

 The proofs are provided in the next
section.

\vspace{0cm plus 4cm}

\section{Proofs}\label{sec:pr}
This section is devoted to the proofs of the results stated in the
last section. In order to prove the Gibbsianness we will first
observe that it is the case for the FRPF's of compact supported
operators. Then we prove that the FRPF's of our concern are weak
limits of such measures. We will apply these facts after
approximating some bounded measurable functions by good bounded
continuous functions.

\subsection{Proof of Proposition \ref{prop:specification}}
In this subsection we prove the construction of Gibbsian
specification, Proposition \ref{prop:specification}. First we
prove Lemma \ref{lem:specification-density}. Recall the notations
$H_\L^{(J)}$ and $\mcR_\L$, respectively in
(\ref{eq:energy-with-b.c.}) and (\ref{eq:possible-event}).

\Proof[ of Lemma \ref{lem:specification-density}] Let $\L\ssbs
\mbR^d$, $\z_\L\in \G_\L$, and $\x\in \mcR_\L$. Without loss, we
consider only the case of the interaction range $R=+\infty$. (For
the case of $R<\infty$, we only need to use $\hat\x$ for $\x$
below.) For any bounded Borel set $\D\sps\L$ define
\begin{equation}
H_{\L;\D}^{(J)}(\z_\L;\x):=U^{(J)}(\z_\L)+W^{(J)}(\z_\L;\x_{\D\sm\L}).
\end{equation}
Since $\x\in \mcR_\L$, $\det J(\x_{\D\sm\L},\x_{\D\sm\L})\neq 0$. That is, $U^{(J)}(\x_{\D\sm\L})<\infty$ for
all bounded $\D\sps\L$. Therefore, by (\ref{eq:mutual_potential}) and (\ref{eq:potential}) we see that
\begin{eqnarray}\label{eq:hamiltonian-reexpression}
H_{\L;\D}^{(J)}(\z_\L;\x)&=&U^{(J)}(\z_\L\x_{\D\sm\L})-U^{(J)}(\x_{\D\sm\L})\nonumber\\
&=&-\log\frac{\det J(\z_\L\x_{\D\sm\L},\z_\L\x_{\D\sm\L})}{\det
J(\x_{\D\sm\L},\x_{\D\sm\L})}.
\end{eqnarray}
If $\det J(\z_\L\x_{\D\sm\L},\z_\L\x_{\D\sm\L})=0 $ for some
$\D\sps\L$, then $\det J(\z_\L\x_{\D'\sm\L},\z_\L\x_{\D'\sm\L})=0$
for all $\D'\sps\D$ (see (\ref{eq:consistency})), and thus we are
done. We suppose $\det J(\z_\L\x_{\D\sm\L},\z_\L\x_{\D\sm\L})\neq
0$ for all $\D\sps\L$ (in particular, $
J(\z_\L\x_{\D\sm\L},\z_\L\x_{\D\sm\L})$ and its submatrices are
invertible). By an elementary manipulation on determinants of
finite matrices we have the identity:
\begin{equation}\label{eq:determinant-identy}
\frac{\det J(\z_\L\x_{\D\sm\L},\z_\L\x_{\D\sm\L})}{\det
J(\x_{\D\sm\L},\x_{\D\sm\L})} =\det (J(\z_\L ,\z_\L)-J(\z_\L,
\x_{\D\sm\L})J( \x_{\D\sm\L}, \x_{\D\sm\L})^{-1}J(
\x_{\D\sm\L},\z_\L)),
\end{equation}
where we have used the obvious notations, e.g.,  $J(\z_\L ,
\x_{\D\sm\L})$ is the matrix
\begin{equation}
J(\z_\L , \x_{\D\sm\L})=(J(x_i,y_j))_{x_i\in \z_\L;\,y_j\in
\x_{\D\sm\L}}.
\end{equation}
 Let $l^2(\z_\L\x_{\L^c})$ be the (complex-valued) $l^2$-space with index set
 $\z_\L\x_{\L^c}$. For any $\D\sps\L$ let $Q_\D$ be the projection operator on
 $l^2(\z_\L\x_{\L^c})$ onto $l^2(\z_\L\x_{\D\sm\L})$. For a convenience we understand
 $J(\z_\L\x_{\D\sm\L},\z_\L\x_{\D\sm\L})$ as $Q_\D
 J(\z_\L\x_{\D\sm\L},\z_\L\x_{\D\sm\L})^\sim
 Q_\D$ where $J(\z_\L\x_{\D\sm\L},\z_\L\x_{\D\sm\L})^\sim$ $:=J(\z_\L\x_{\D\sm\L},
 \z_\L\x_{\D\sm\L})\oplus
 \bf{1}$  is a bounded linear operator on $l^2(\z_\L\x_{\L^c})\equiv l^2(\z_\L\x_{\D\sm\L})\oplus
 l^2(\x_{\D^c})$ acting as an identity operator on $l^2(\x_{\D^c})$. We notice that the r.h.s.
 of (\ref{eq:determinant-identy}) is equal to
 \begin{equation}\label{eq:equivalent-form}
 \big(\det (Q_\L J(\z_\L\x_{\D\sm\L},\z_\L\x_{\D\sm\L})^{-1}Q_\L)\big)^{-1}.
 \end{equation}
 On the other hand, for any bounded operator $T$ with bounded
 inverse $T^{-1}$ and any projection $P$, from
 the decomposition (see \cite[page 18]{OP} and \cite{GY} for a proof)
 \begin{equation}\label{eq:docomposition}
PT^{-1}P=P(PTP)^{-1}P+PT^{-1}P^{\bot}(P^{\bot}T^{-1}P^{\bot})^{-1}P^{\bot}T^{-1}P,
\end{equation}
we get the inequality
\begin{equation}\label{eq:basic-order}
PT^{-1}P\ge P(PTP)^{-1}P.
\end{equation}
Therefore if $\D'\sps\D$, then by replacing $P=Q_\D$ and
$T=J(\z_\L\x_{\D'\sm\L},\z_\L\x_{\D'\sm\L})$ in
(\ref{eq:basic-order}) we get
\begin{eqnarray}\label{eq:monotonicity1}
Q_\D J(\z_\L\x_{\D'\sm\L},\z_\L\x_{\D'\sm\L})^{-1}Q_\D
&\ge&Q_\D(Q_\D
J(\z_\L\x_{\D'\sm\L},\z_\L\x_{\D'\sm\L})Q_\D)^{-1}Q_\D\nonumber\\
&=&Q_\D J(\z_\L\x_{\D\sm\L},\z_\L\x_{\D\sm\L})^{-1}Q_\D\nonumber\\
&=&J(\z_\L\x_{\D\sm\L},\z_\L\x_{\D\sm\L})^{-1}.
\end{eqnarray}
Applying $Q_\L$ from left and right of both sides of
(\ref{eq:monotonicity1}) we see that
\begin{equation}\label{eq:monotonicity2}
Q_\L J(\z_\L\x_{\D'\sm\L},\z_\L\x_{\D'\sm\L})^{-1}Q_\L\ge Q_\L
J(\z_\L\x_{\D\sm\L},\z_\L\x_{\D\sm\L})^{-1}Q_\L,\quad \D'\sps\D,
\end{equation}
as operators on $l^2(\z_\L)$. Notice also that if $0\le A\le B$
are two positive definite $n\times n$  matrices then
\begin{equation}
0\le \l_i^{\da}(A)\le \l_i^{\da}(B), \quad i=1,\cdots,n,
\end{equation}
where $\l_i^{\da}(A)$ ($\l_i^{\da}(B)$, respectively),
$i=1,\cdots, n$, are the eigenvalues of $A$ (of $B$, respectively)
ordered in decreasing order \cite[Corollary III.2.3]{B}. Therefore
from (\ref{eq:determinant-identy}), (\ref{eq:equivalent-form}),
and (\ref{eq:monotonicity2}), the l.h.s. of
(\ref{eq:determinant-identy}) decreases as $\D$ increases. From
this and monotonicity of logarithmic function it follows that the
limit
\begin{equation}\label{eq:existence}
H_\L^{(J)}(\z_\L;\x):=\lim_{\D\ua \mbR^d}H_{\L;\D}^{(J)}(\z_\L;\x)
\end{equation}
exists. \EndProof

Now let us recall the definition of $\g_\L^{(J)}(\z_\L;\x)$ in
(\ref{eq:specification-density}). For $\x\in \mcR_\L$, the
partition function $Z_\L^{(J)}(\x)$ is defined by
\begin{equation}\label{eq:partition-function}
Z_\L^{(J)}(\x)=1+\sum_{n\ge 1}\frac1{n!}\int_{\L^n}d\ul
x_n\exp[-H_\L^{(J)}(\ul x_n;\x)].
\end{equation}
In the case $R=\infty$, from (\ref{eq:existence}) and
(\ref{eq:hamiltonian-reexpression})-(\ref{eq:determinant-identy})
we see that
\begin{eqnarray}\label{eq:boltzman-factor-bound}
\exp[-H_\L^{(J)}(\ul x_n;\x)]&=&\lim_{\D\ua
\mbR^d}\exp[-H_{\L;\D}^{(J)}(\ul
x_n;\x)]\nonumber\\
&=&\lim_{\D\ua \mbR^d}\det (J(\ul x_n ,\ul x_n)-J(\ul x_n,
\x_{\D\sm\L})J( \x_{\D\sm\L},
\x_{\D\sm\L})^{-1}J( \x_{\D\sm\L},\ul x_n))\nonumber\\
&\le &\det J(\ul x_n,\ul x_n).
\end{eqnarray}
Thus from
(\ref{eq:partition-function})-(\ref{eq:boltzman-factor-bound}) we
have
\begin{equation}\label{eq:partition-function-bound}
1\le Z_\L^{(J)}(\x)\le 1+\sum_{n\ge 1}\frac1{n!}\int_{\L^n}d\ul
x_n \det J(\ul x_n,\ul x_n)=\det (I+J_\L).
\end{equation}
In the last equality we have used a formula for the Fredholm
determinant \cite[Theorem 3.10]{Sim}. In the case $R<\infty$, we
just replace $\x$ by $\hat\x$ (see \eqref{eq:energy-with-b.c.}) in
the above, and we get (\ref{eq:partition-function-bound}), too.

We now prove Proposition \ref{prop:specification}.

\Proof[ of Proposition \ref{prop:specification}] Let us define for
$\L\ssbs\mbR^d$, $A\in \mcF$, and $\x\in \mcN$,
\begin{equation}\label{eq:specification-kernel}
\g_\L^{(J)}(A|\x):=\g_\L^{(J)}(1_A|\x),
\end{equation}
where $1_A$ is the characteristic function on the set $A$ (see
(\ref{eq:specification})). We have to show that (see \cite[page
16]{Pr}):

(i) $\g_\L^{(J)}(\cdot|\x)$ is a probability measure for each
$\x\in {\mcR}_\L$, $\L\ssbs \mbR^d$;

(ii) $\g_\L^{(J)}(A|\x)=0$ for all $\x\notin {\mcR}_\L$,
$\L\ssbs\mbR^d$, $A\in \mcF$;

(iii) $\g_\L^{(J)}(A|\cdot)$ is $\mcF_{\L^c}$-measurable for all
$A\in \mcF$, $\L\ssbs\mbR^d$;

(iv) $\g_\L^{(J)}(A|\cdot)=1_{A\cap {\mcR}_\L}$ if $A\in
\mcF_{\L^c}$, $\L\ssbs \mbR^d$;

(v)
$\g_\D^{(J)}\g_\L^{(J)}(A|\x):=\int\g_\D^{(J)}(d\z|\x)\g_\L^{(J)}(A|\z)=\g_\D^{(J)}(A|\x)$
whenever $\L\sbs\D\ssbs\mbR^d$.

\noindent From the definition, the properties (i)-(iv) are
obvious. The proof of (v) is a routine, but a simple computation
by noticing the product property of the measure:
\begin{equation}
\oint_\D d\z_\D f(\z_\D)=\oint_\L d\z_\L \oint_{\D\sm\L}
d\z_{\D\sm\L} f(\z_\L\z_{\D\sm\L}),\quad \L\sbs\D,
\end{equation}
which is easily shown as follows:
\begin{eqnarray*}
&&\oint_\L d\z_\L \oint_{\D\sm\L} d\z_{\D\sm\L}
f(\z_\L\z_{\D\sm\L})\\
&=&\sum_{l\ge 0}\frac1{l!}\int_{\L^l}d\ul{x}_l\sum_{m\ge
0}\frac1{m!}\int_{(\D\sm\L)^m}d\ul{y}_m\,f(\ul{x}_l\ul{y}_m)\\
&=&\sum_{n\ge 0}\frac1{n!}\int_{\D^n}d\ul{z}_n\,f(\ul{z}_n)\\
&=&\oint_\D d\z_\D f(\z_\D).
\end{eqnarray*}
In the second equality, we have put $l+m=n$ and
$\ul{x}_l\ul{y}_m=\ul{z}_n$. The proof is completed. \EndProof

\subsection{Proof of Gibbsianness}
In order to prove Theorem~\ref{thm:Gibbsian} we need some
preparations. First we will observe that FRPF's corresponding to
compact supported operators are Gibbs measures.

Let $J$ be an integral operator supported on a compact region in
$\mbR^d$. That is, there is a compact set $\D_0\subset\mbR^d$ such
that the kernel function $J(x,y)$ satisfies $J(x,y)=0$ unless $x$
and $y$ belong to $\D_0$. We define $\g _{\L }^{(J)}(\ul x_n;\x )$
as in (\ref{eq:specification-density}) (with a suitable $\mcR_\L$,
e.g., $\mcR_\L:=\{\x:\,\x_{(\L\cup\D_0)^c}=\emptyset\}$). It is
obvious that $\g _{\L }^{(J)}(\ul x_n;\x )=0$ whenever $\x
_{\D_0^c}\neq \emptyset$ and it is well-defined for all (finite)
configurations $\x $ in $\D_0$. Using this density function we
define a specification $(\g _{\L }^{(J)})_{\L \ssbs\mbR^d}$
through the formula (\ref{eq:specification}). Let $\mu^{(J)} $ be
the FRPF corresponding to the operator $K:=J(I+J)^{-1}$. We notice
that the operator $K$ is also supported on $\D_0$ and hence
$\mu^{(J)}$ is supported on the set $ \G_{\D_0}$. The following is
a key observation:
\begin{prop}\label{prop:key_observation}
Let $J$ be an integral operator with compact support. Define a
specification $(\g _{\L }^{(J)})_{\L \ssbs\mbR^d}$ and a FRPF
$\mu^{(J)}$ as above. Then $\mu^{(J)}$ is a Gibbs measure for $(\g
_{\L }^{(J)})_{\L \ssbs\mbR^d}$.
\end{prop}
\Proof For simplicity we omit all the superscripts from the
notations. Let $A\in \mcF$ and $\L \ssbs \mbR^d$. We have to show
that
\begin{equation}
\mu(A)=\int d\mu(\x )\g _\L (A|\x ). \label{eq:Gibbs_property}
\end{equation}
Since $\g _\L (A|\x )=0$ if $\xi_{\D_0^c}\neq \emptyset$, by using
(\ref{eq:density_function}), (\ref{eq:specification-density}), and
(\ref{eq:specification-kernel}) the r.h.s. equals to
\begin{eqnarray}
  &&\oint_{\D_0}d\x _{\D_0}\sigma_{\D_0}(\x _{\D_0})\g _\L (A|\x _{\D_0}) \nonumber\\
&=&\det(I-K)\oint_{\D_0}d\x _{\D_0} \det J(\x _{\D_0},\x
_{\D_0})\nonumber\\
&&\quad\quad\times \frac1{Z_\L(\x_{\D_0})}\oint_{\L }d\z _{\L }
\frac{\det J(\z _\L \x _{\D_0\sm\L },\z _\L \x _{\D_0\sm\L })}
{\det J(\x_{\D_0\sm\L},\x_{\D_0\sm\L})}
1_A(\z _\L \x _{\D_0\sm\L })\nonumber\\
&=&\det(I-K)\oint_{\D_0\sm\L}d\x_{\D_0\sm\L}\frac1{Z_\L(\x_{\D_0})}\oint_{\L
}d\x _{\L } \frac{\det J(\x _{\D_0},\x _{\D_0})}{\det
J(\x_{\D_0\sm\L},\x_{\D_0\sm\L})}\nonumber\\
&&\quad\quad\times\oint_{\L }d\z _{\L } \det J(\z _\L \x
_{\D_0\sm\L },\z _\L \x _{\D_0\sm\L }) 1_A(\z _\L
\x_{\D_0\sm\L })\nonumber\\
 &=&\det(I-K)\oint_{\D_0\sm\L }d\x _{\D_0\sm\L
}\oint_{\L }d\z _{\L }  \det J(\z _\L\x_{\D_0\sm\L},\z_\L \x
_{\D_0\sm\L })1_A(\z _\L \x _{\D_0\sm\L })\nonumber\\
&=&\mu(A).
\end{eqnarray}
In the first and last equalities we have used $K_{\D_0}=K$ and
$J_{[\D_0]}=K_{\D_0}(I-K_{\D_0})^{-1}=J$. The fractions are set to
be zero if the denominator equals to zero by the property
(\ref{eq:consistency}). We have proven (\ref{eq:Gibbs_property}).
\EndProof

Next we discuss some weak convergence of FRPF's.  The following
may be well known.
\begin{lem}\label{lem:trace_norm_convergence}
Let $J$ be a bounded, positive definite, locally trace class
integral operator. Let $(\L _n)_{n\ge 1}$ be any increasing
sequence of bounded Borel subsets of $\mbR^d$ such that $\cup_n\L
_n=\mbR^d$. Define $K_n:=J_{\L _n}(I+J_{\L _n})^{-1}$. Then for
any $\L \ssbs\mbR^d$, $\|P_\L  K_n P_\L -P_\L  K P_\L \|_1\to 0$
as $n\to \infty$, where $K:=J(I+J)^{-1}$.
\end{lem}
\Proof Let $B:=P_{\L }JP_{\L }$. Then $B$ is a trace class
operator and $P_{\L }KP_{\L }\le B$ and $P_{\L }K_nP_{\L }\le
P_{\L }J_{\L _n}P_{\L }=B$ whenever $\L _n\supset \L $. Moreover,
$P_{\L }K_nP_{\L }\to P_{\L }KP_{\L }$ weakly since $P_{\L _n}$
converges strongly to the identy. Now we use Theorem 2.16 of
\cite{Sim} to complete the proof. \EndProof

\begin{lem}\label{lem:weak_convergence} (cf. \cite{ST1, So}) We suppose the same
setting as in Lemma \ref{lem:trace_norm_convergence}. Then   the
sequence of FRPF's $\m_n^{(J)}$ corresponding to $K_n $ converges
weakly to the FRPF $\mu^{(J )}$ corresponding to $K$.
\end{lem}
\Proof By using Lemma \ref{lem:trace_norm_convergence} the proof
follows from \cite[Theorem 5]{So}. We provide however a proof
here. We will show that for any bounded, measurable, and local
functions $F:\G\to \mbR$ (we emphasize again that FRPF's are
supported on $\G$),
\begin{equation}\label{eq:convergence}
\int F(\x)d\m_n^{(J)}(\x)\to \int F(\x)d\m^{(J)}(\x) \quad
\text{as }n\to \infty.
\end{equation}
For any $\L\ssbs\mbR^d$, $\G_\L$ is isomorphic to the space of
disjoint sum $\sum_{n\ge 0}\widetilde{\L^n}$, where
$\widetilde{\L^n}$ is the symmetric space of $n$ different points
in $\L$ \cite{DV}. In particular, it is a locally compact space.
By Stone-Weierstrass theorem it is therefore enough to show
(\ref{eq:convergence}) only for $F\in \mcC_{+}$, where $\mcC_{+}$
is defined by
\begin{equation}
\mcC_{+}:=\{F:\G\to \mbR:\,F(\x)=e^{-<f,\x>}\text{ for some }0\le
f\in C_0(\mbR^d)\}.
\end{equation}
For such an $F(\x)=e^{-<f,\x>}$, we have by
(\ref{eq:Laplace_transform})
\begin{equation}
\int d\mu_n^{(J )}(\x )e^{-<f,\x >}=\det (I- K_n  \psi_f).
\label{eq:LT_of_K_n}
\end{equation}
Since $0\le f  \in C_0(\mbR^d)$, we have also $0\le \psi_f\in
C_0(\mbR^d)$. Thus we can rewrite
\begin{equation}\label{eq:symmetric_expression}
\det(I-K_n\psi_f)=\det(I-\sqrt {\psi_f} K_n\sqrt{\psi_f}).
\end{equation}
 By Lemma~\ref{lem:trace_norm_convergence}, $\sqrt {\psi_f} K_n\sqrt{\psi_f} $ converges to
 $ \sqrt {\psi_f} K \sqrt{\psi_f}$ in trace norm. Since $A\mapsto \det (I+A)$ is continuous
 in trace
 norm (\cite[Theorem 3.4]{Sim}), the r.h.s. of (\ref{eq:symmetric_expression}) converges to
\begin{equation}
\det (I- \sqrt {\psi_f} K \sqrt{\psi_f})=\int d\mu^{(J )}(\x
)e^{-<f,\x  >}.
\end{equation}
By \cite[Proposition 9.1.VII]{DV}, (\ref{eq:convergence}) is
already equivalent to the weak convergence. The proof is
completed. \EndProof

In order to prove the Gibbsianness of our $\m^{(J)}$ in Theorem
\ref{thm:Gibbsian} we will use Proposition
\ref{prop:key_observation} and Lemma \ref{lem:weak_convergence}.
First notice that instead of showing (\ref{eq:Gibbs_property}) it
is enough to show that for all $\L\ssbs\mbR^d$ and any bounded and
continuous function $f$
\begin{equation}\label{eq:Gibbs-property-function-version}
\int d\m^{(J)}(\x)f(\x)=\int d\m^{(J)}(\x)\g_\L^{(J)}(f|\x).
\end{equation}
Let $\m_n^{(J)}$ be the FRPF's weakly converging to $\m^{(J)}$ as
in Lemma \ref{lem:weak_convergence}. It turns out that
(\ref{eq:Gibbs-property-function-version}) holds true for
$\m_n^{(J)}$'s. For our goal we want to let $n$ tend to infinity.
The difficulty in this step occurs because we do not know the
continuity of the function $\x\mapsto \g_\L^{(J)}(f|\x)$ in
general. To overcome this difficulty we need to approximate the
function $\g_\L^{(J)}(f|\x)$ by good continuous functions. For
that purpose we rely on finiteness of the range and the
non-percolating property of FRPF's for small intensity.

Let $\x=(x_i)_{i=1,2,\cdots}\in \mcN$ be a configuration. Let
$R>0$ be the number in Assumption \ref{ass:assumption2}. For each
$i=1,2,\cdots$, we position a closed $d$-dimensional sphere $S_i$
of fixed radius $R$ with center $x_i$.  We call two spheres $S_i$
and $S_j$ adjacent if $S_i\cap S_j\neq \emptyset$. We write
$S_i\leftrightarrow S_j$ if there exists a sequence $S_{i_1},
S_{i_2}, \cdots, S_{i_k}$ of spheres such that $S_{i_1}=S_i$,
$S_{i_k}=S_j$, and $S_{i_l}$ is adjacent to $S_{i_{l+1}}$ for
$1\le l<k$. A cluster of spheres is a set $(S_i:i\in I)$ of
spheres which is maximal with the property that
$S_i\leftrightarrow S_j$ for all $i,j\in I$. The size of a cluster
is the number of spheres belonging to it. The following was proved
in \cite[Corollary 3.5]{GY}:
\begin{thm}\label{thm:no-percolation}
Suppose that $J$ satisfies the conditions in Assumption \ref{ass:assumption1}. Then, there is a critical
intensity $\a_c(d)>0$ ($\a_c(1)=\infty$, in particular) such that if $j(0)<\a_c(d)$, then
\begin{equation}\label{eq:no-cluster}
\m^{(J)}(\text{there is an infinite cluster})=0.
\end{equation}
\end{thm}
We are now ready to prove the main result.

\Proof[ of Theorem \ref{thm:Gibbsian}] Let $f:\mcN\to \mbR$ be any
bounded and continuous function and $\L\ssbs \mbR^d$. We have to
show (\ref{eq:Gibbs-property-function-version}). Let $(\L
_n)_{n\ge 1}$ be any increasing sequence of bounded Borel subsets
of $\mbR^d$ with $\cup_n\L _n=\mbR^d$. We define $K_n := J_{\L
_n}({I+ J_{\L _n}})^{-1}$ and $J_n :=K_n (I-K_n )^{-1}= J_{\L
_n}$. Let $\m_n^{(J)}$ be the FRPF corresponding to $K_n$. Since
$J_n $ is compactly supported we have by
Proposition~\ref{prop:key_observation}
\begin{equation}
\mu_n^{(J )}(f)=\int d\mu_n^{(J )}(\x )\g _\L ^{(J_n )}(f|\x ).
\label{eq:main_n_Gibbs_property}
\end{equation}
We recall that $\g _\L ^{(J_n )}(f|\x )=0$ if $\x _{\L _n^c}\neq
\emptyset$ and $\mu_n^{(J )}$ is supported on $\G_{\L_n}$. On the
other hand, we see that for $\x =\x _{\L _n}$ (i.e., $\x _{\L
_n^c}= \emptyset$)
\begin{eqnarray*}
\g _\L ^{(J_n )}(f|\x)=\g _\L ^{(J_n )}(f|\x _{\L _n})
&=&\frac{1}{Z_{\L} ^{(J_n )} (\x_{\L_n})} \oint _\L d\z _\L
\frac{\det  J_{\L _n}(\z _\L \x _{\L _n\sm\L },\z _\L \x _{\L
_n\sm\L })}{\det J_{\L _n}(\x _{\L _n\sm\L },\x _{\L _n\sm\L })}
f(\z _\L\x_{\L_n\sm\L})\\
&=&\g_\L^{(J )}(f|\x_{\L_n})=\g_\L^{(J )}(f|\x).
\end{eqnarray*}
In the third equality we have used the fact that $J_{\L _n}(\et
_{\L _n },\et_{\L _n })=J(\et _{\L _n },\et_{\L _n })$ for any
$\et_{\L_n}\in \G_{\L_n}$. Thus (\ref{eq:main_n_Gibbs_property})
is equivalent to the equation
\begin{equation}\label{eq:gibbsian-for-mu_n}
\mu_n^{(J )}(f)=\int d\mu_n^{(J )}(\x )\g _\L ^{(J )}(f|\x ).
\end{equation}

For $d\ge 2$, we assume that $j(0)$ is sufficiently small that
(\ref{eq:no-cluster}) holds. We let
\begin{equation}\label{eq:configurations-of-infinite-cluster}
\mcC_\L:=\{\x\in \mcN:\,\x_{\L^c} \text{ has an infinite cluster
connected to }\L\}
\end{equation}
and define
\begin{equation}\label{eq:an-open-set}
\mcO_\L:=\mcR_\L\cap\mcC_\L^c.
\end{equation}
A short consideration tells us that $\mcO_\L$ is an
$\mcF_{\L^c}$-measurable, open subset of $\mcN$. In Lemma
\ref{lem:full_charger}  below we will show that
\begin{equation}\label{eq:full-charger}
\m^{(J)}(\mcO_\L)=1.
\end{equation}
We also consider the clusters of open balls centered at the
particles.  Let $B_0$ be a big enough closed ball in $\mbR^d$
centered at the origin and such that $B_0\sps\L$ and
$\text{dist}(\L,B_0^c)>2R+1$, where $\text{dist}(\L,B_0^c)$
denotes the distance between $\L$ and $B_0^c$. Let $B_1\sbs
B_2\sbs\cdots$ be any increasing sequence of closed balls in
$\mbR^d$ centered at the origin such that $B_0\sbs B_1$ and
$\cup_n B_n=\mbR^d$. For each $n=1,2,\cdots$, we let $R_n:=R+1/n$
and define
\begin{equation}\label{eq:configurations-of-no-cluster}
\mcC_\L^{(n)}:=\{\x\in \mcN:\,\text{ clusters } (\text{of } \x)
\text{ of }R_n\text{-open balls connected to } B_0 \text{ do not
reach }B_n^c\}.
\end{equation}
We let $\mcD_\L\equiv \mcO_\L^c$ and
\begin{equation}
\mcD_\L^{(n)}:=\{\x\in \mcN:\,\x\in \mcC_\L^{(n)},\,\,\det
J(\x_{\ptl\L},\x_{\ptl\L})\ge 1/n\}.
\end{equation}
We notice that $(\mcD_\L^{(n)})_{n\ge 1}$ is a sequence of
increasing closed subsets of $\mcN$ and it is not hard to see that
\begin{equation}\label{eq:full-of-no-cluster-configurations}
\cup_{n\ge 1}\mcD_\L^{(n)}=\mcO_\L.
\end{equation}
Let $d$ be a metric that makes $\mcN$ a complete separable metric
space and the metric topology is equivalent to the vague topology.
For each $n=1,2,\cdots$, define a function $\ch_n:\mcN\to [0,1]$
by
\begin{equation}\label{eq:approximation}
\ch_n(\x):=\frac{d(\x,\mcD_\L)}{d(\x,\mcD_\L^{(n)})+d(\x,\mcD_\L)}.
\end{equation}
We notice that $\ch_n$ is continuous and $\ch_n=1$ on
$\mcD_\L^{(n)}$ and $\ch_n=0$ on $\mcD_\L$, and by
(\ref{eq:full-charger}) and
(\ref{eq:full-of-no-cluster-configurations})
\begin{equation}\label{eq:approximation2}
\lim_{n\to \infty}\ch_n\nearrow 1\quad\m^{(J)}\text{-a.e.}
\end{equation}
Moreover, since $\mcD_\L^{(n)}$ and $\mcD_\L$ are both
$\mcF_{\L^c}$-measurable, the function $\ch_n$ is also
$\mcF_{\L^c}$-measurable. Observe that for each fixed
$n=1,2,\cdots$ the function $\x\mapsto \g_\L^{(J)}(\ch_nf|\x)$ is
continuous. In fact, since $\ch_n$ is $\mcF_{\L^c}$-measurable, we
have
\begin{eqnarray}\label{eq:continuity}
\g_\L^{(J)}(\ch_nf|\x)&=&\ch_n(\x)\g_\L^{(J)}(f|\x)\nonumber\\
&=&\ch_n(\x)\oint_\L d\z_\L\g_\L^{(J)}(\z_\L;\x)f(\z_\L\x_{\L^c}).
\end{eqnarray}
For $\x\in \mcD_\L$, we have $\g_\L^{(J)}(\ch_nf|\x)=0$ and
$|\g_\L^{(J)}(\ch_nf|\et)|\le \ch_n(\et)\|f\|_{\infty}$, $\et\in
\mcN$. Thus $\g_\L^{(J)}(\ch_nf|\cdot)$ is continuous at $\x$. On
the other hand, if $\x\in \mcO_\L$, then there is no infinite
cluster (of closed $R$-spheres) of $\x_{\L^c}$ connected to $\L$
and therefore we have by definition
\begin{equation}\label{eq:localized-function}
\g_\L^{(J)}(\z_\L;\x)=\g_\L^{(J)}(\z_\L;\x_{\ptl\L}).
\end{equation}
Furthermore, since $\det J(\x_{\ptl\L}, \x_{\ptl\L})>0$, we can
find some bounded and, say, open set $\D_0\sbs\mbR^d$ and an open
neighborhood $U$ of $\x$ such that for each $\et\in U$,
$\et_{\ptl\L}=\et_{\D_0\sm\L}$ and $\det
J(\et_{\ptl\L},\et_{\ptl\L})>0$, i.e.,
\begin{equation}\label{eq:continuity-out-of-C}
\g_\L^{(J)}(\z_\L;\et)=\g_\L^{(J)}(\z_\L;\et_{\D_0})\text{ and
}\det J(\et_{\D_0\sm\L},\et_{\D_0\sm\L})>0, \quad \et\in U.
\end{equation}
Since the function $\et\mapsto\g_\L^{(J)}(\z_\L;\et_{\D_0})$ is
continuous on $U$, the function
 $\et\mapsto\g_\L^{(J)}(\z_\L|\et)$ is continuous at $\x$ and therefore so is the function
 $\g_\L^{(J)}(\ch_nf|\cdot)$. From (\ref{eq:gibbsian-for-mu_n}) we have
\begin{equation}\label{eq:gibbsian-for-mu_n-with-approximation}
\mu_m^{(J )}(\ch_nf)=\int d\mu_m^{(J )}(\x )\g _\L ^{(J
)}(\ch_nf|\x ).
\end{equation}
Now both $\ch_nf$ and $\g_\L^{(J)}(\ch_nf|\cdot)$ are bounded and
continuous and $\m_m^{(J)}$ converges weakly to $\m^{(J)}$ as $m$
tends to infinity. By letting $m$ go to infinity in
(\ref{eq:gibbsian-for-mu_n-with-approximation}) we get
\begin{eqnarray}\label{eq:gibbsian-for-approximated-functions}
\mu^{(J )}(\ch_nf)&=&\int d\mu^{(J )}(\x )\g _\L ^{(J )}(\ch_nf|\x )\nonumber\\
&=&\int d\mu^{(J )}(\x )\ch_n(\x)\g _\L ^{(J )}(f|\x ).
\end{eqnarray}
We let now $n$ tend to infinity and use (\ref{eq:approximation2})
to get
\begin{equation}
\mu^{(J )}(f)=\int d\mu^{(J )}(\x )\g _\L ^{(J )}(f|\x ).
\end{equation}
The proof is completed. \EndProof

Finally, we provide with a proof \eqref{eq:full-charger}. For that
purpose we need the following property.
\begin{lem}\label{lem:continuous_kernel}
Let $J$ be an operator satisfying the conditions in Assumption
\ref{ass:assumption1} with $d\r(t)=\hat{\ph}(t)dt$ for some $0\le
\hat{\ph}(\cdot)\in L^1(\mbR^d)$. Then for each $\L\ssbs\mbR^d$,
the operator $J_{[\L]}$ admits a continuous kernel function
$J_{[\L]}(x,y)$, $x,y\in \L$, where $J_{[\L]}:=K_\L(I-K_\L)^{-1}$
with $K:=J(I+J)^{-1}$.
\end{lem}
\Proof It is readily seen that $K$ has an integral kernel function
$K(x,y)=k(x-y)$, where
\begin{equation}\label{eq:k_kernel}
k(x):=(2\p)^{-d}\int_{\mbR^d}e^{ix\cdot
t}\frac{\hat{\ph}(t)}{1+\hat{\ph}(t)}dt.
\end{equation}
In particular, $k(\cdot)$ is continuous. Now let $J_{[\L]}'(x,y)$,
$x,y\in \L$, be any kernel function of the operator $J_{[\L]}$,
which exists because $J_{[\L]}$ is a Hilbert-Schmidt operator.
Since
\[J_{[\L]}=K_\L+(K_\L)^2+K_\L J_{[\L]}K_\L,\]
$J_{[\L]}'(x,y)$ coincides for almost all $(x,y)\in\L^2$ with
\[\begin{split}
J_{[\L]}(x,y):= & \; k(x-y) + \int_\L k(x-u)\,k(u-y)\,du\\ & +
\int_\L\int_\L k(x-u)\,J_{[\L]}'(u,v)\,k(v-y)\,du\, dv\,.
\end{split}
\]
Since $k$ is continuous, it is easily checked that $J_{[\L]}(x,y)$
is continuous, as required.\EndProof

\begin{lem}\label{lem:full_charger}
For each $\L\ssbs \mbR^d$, let $\mcO_\L$ be defined as in
(\ref{eq:an-open-set}). Then we have
\begin{equation}\label{eq:full-charger1}
\m^{(J)}(\mcO_\L)=1.
\end{equation}
\end{lem}
\Proof Since $\m^{(J)}(\mcC_\L)=0$ by (\ref{eq:no-cluster}), it is
enough to show that
\begin{equation}\label{eq:enough-set}
\m^{(J)}(\mcR_\L^c\cap \mcC_\L^c)=0.
\end{equation}
We let $\mcG$ be a countable class of bounded open subsets of
$\mbR^d$ such that for any compact subset $C\sbs\mbR^d$ and
$z\notin C$, we can find a $G\in \mcG$ such that
\begin{equation}
C\sbs G \text{ and }z\notin G.
\end{equation}
It is obvious that
\begin{equation}\label{eq:an-inclusion}
\mcR_\L^c\cap \mcC_\L^c\sbs \cup_{G\in \mcG}\cup_{n\ge
1}\mcN_{G,n},
\end{equation}
where
\begin{equation}\label{eq:a-local-set}
\mcN_{G,n}:=\{\x\in \mcN:\,\x(G\sm\L)=n \text{ and } \det
J(\x_{G\sm\L},\x_{G\sm\L})=0\}.
\end{equation}
On the other hand, by using the inequality \eqref{eq:basic-order},
it is not hard to show the operator ordering: $J_{[\D]}\le J_\D$
for all $\D\ssbs\mbR^d$ (see \cite[Lemma 4.1]{GY} for a proof).
Moreover, since $J(x,y)$ is continuous and $J_{[\D]}$ also admits
a continuous kernel $J_{[\D]}(x,y)$ by Lemma
\ref{lem:continuous_kernel}, we have for any $\D\ssbs\mbR^d$ and
$n\ge 1$, $(J_{[\D]}(x_i,x_j))_{1\le i,j\le j}\le
(J(x_i,x_j))_{1\le i,j\le j}$ as operators on $\mbC^n$, and thus
\begin{equation}\label{eq:a-comparison}
\det(J_{[\D]}(x_i,x_j))_{1\le i,j\le j}\le \det(J(x_i,x_j))_{1\le
i,j\le j}
\end{equation}
for all $(x_1,\cdots,x_n)\in \D^n$. We therefore get by
(\ref{eq:density_function}), (\ref{eq:a-local-set}), and
(\ref{eq:a-comparison}),
\begin{equation}\label{eq:no-charge}
\m^{(J)}(\mcN_{G,n})=0,\quad \text{for all }G\in \mcG,\,\,n\ge 1.
\end{equation}
By (\ref{eq:an-inclusion}) and (\ref{eq:no-charge}) we get
(\ref{eq:enough-set}) and the proof is complete. \EndProof

\vspace{0 cm plus 4cm}

\section{Concluding remarks}\label{sec:conc}
In this section we would like to discuss some possible
improvements of the results obtained in this paper, which were
done in \cite{GY}.

First, though the typical examples of operators $J$ might be the
ones given in Assumptions \ref{ass:assumption1} and
\ref{ass:assumption2}, from the pedagogical point of view, the
class of operators $J$ (and hence the operators $K$) that have the
properties in the main results of this paper is in fact much
larger than the one we considered. It is possible to include, for
example, the operators with non-continuous kernel functions and
the ones that are not of translation-invariant. To extend the
theory to that generality, we are, however, confronted with some
subtlety of version-problems of the kernel function. This was
thoroughly investigated in \cite{GY}.

Second, as remarked in the paragraph after the proof of Lemma \ref{lem:weak_convergence}, the key idea to show
the Gibbsianness is to get a continuity of the function $\x\mapsto \g_\L(f|\x)$ for any continuous function $f$
with local support. The condition of finite range and small intensity was in fact introduced to guarantee such
continuity. Some models, however, possess the continuity without the finiteness condition of the range. An
example is the renewal process on the real line \cite{DV, M}, for which the function $k(\cdot)$ in
\eqref{eq:k_kernel} is given by
\[
k(x):=\r e^{-a|x|},
\]
where $\r,\,a>0$ and $\r<a/2$. For the details, we refer to
\cite[Example 3.11]{GY}.

The most stringent condition is the boundedness of the operator
$J\ge 0$, or the strictness in the ordering $K\le I$ for the
operator $K=J(I+J)^{-1}$. From this restriction, we have to
exclude the most interesting models, for example, the Dyson's
model \cite{Sp}, where the defining operator $K$ has a sine kernel
and $1$ belongs to its spectrum. For those models, the operator
$J:=K(I-K)^{-1}$ is not even defined. So, one asks whether FRPF's
can still be Gibbs measures, and in this case, what the
corresponding interactions are.

\vspace{0cm plus 4cm}
\appendix
\section{Appendix}\label{sec:app}

 In this appendix we provide with some examples
satisfying the conditions in Assumption \ref{ass:assumption2}. Let
$a:\mbR\to \mbR$ be a function defined by
\begin{equation}\label{eq:localizing-function}
a(x):=\begin{cases}-\frac1{R}(x-R),\quad 0\le x<R\\
\frac1{R}(x+R), \quad -R<x<0\\
0,\quad |x|\ge R.
\end{cases}
\end{equation}
Then its Fourier transform is
\begin{eqnarray}
\hat{a}(t)&:=&\int e^{-ixt} a(x)dx\nonumber\\
&=&\frac2{Rt^2}(1-\cos Rt)\ge 0.
\end{eqnarray}
For $x=(x^1,\cdots,x^d)\in \mbR^d$, let
\begin{equation}
u(x):=\prod_{l=1}^d a(x^l).
\end{equation}
Then the Fourier transform is
\begin{equation}
\hat{u}(t)=\prod_{l=1}^d \hat{a}(t^l),\quad t=(t^1,\cdots,t^d)\in
\mbR^d.
\end{equation}
Now let $\hat{\ph}(t)\ge 0$, $t\in \mbR^d$, be any function such
that
\begin{equation}
\int_{\mbR^d} \hat{\ph}(t)\,dt<\infty
\end{equation}
and that its inverse Fourier transform $\ph(x)$ defines a kernel
function of a bounded linear integral operator on $L^2(\mbR^d)$.
Define
\begin{equation}
j(x):=\ph(x)u(x).
\end{equation}
Then
\begin{equation}
\hat{j}(t)=\frac1{2\p}\hat{\ph}\ast \hat{u}(t)=\frac1{2\p}\int
\hat{\ph}(t-s)\hat{u}(s)ds\ge 0
\end{equation}
and
\begin{equation}
\int \hat{j}(t)dt<\infty.
\end{equation}
Therefore, $\hat{j}(t)dt$ is a finite measure on $\mbR^d$. By
Bochner's theorem, its inverse Frourier transform $j(x)$ is a
bounded and continuous function of positive type \cite{RS}.
Obviously we have
\begin{equation}
j(x)=0 \text{ if }|x|\ge R.
\end{equation}
\vskip 0.5 true cm
 \noindent{\it Acknowledgments.} The author is grateful to Prof. M. R\"ockner
 for pointing out some gaps in the first version. He would like to thank
 Prof. Y. Nagahata, Prof. T. Shirai, Prof. Y. M.
Park, and Dr. C. Bahn for fruitful discussions. He wishes also to
thank Prof. T. Shiga for warm hospitality during the stay at Tokyo
Institute of Technology. This work was partially supported by
Japan Society for the Promotion of Science, Research Fellowships,
and by Korea Research Foundation Grant (KRF-2002-015-CP0038).

\end{document}